# Control System for the LEDA 6.7-MeV Proton Beam Halo Experiment[1]

L. A. Day, M. Pieck, D. Barr, K. U. Kasemir, B. A. Quintana, G. A. Salazar, M. W. Stettler
Los Alamos, Los Alamos National Laboratory, NM 87545, USA

*Abstract*

Measurement of high-power proton beam-halo formation is the ongoing scientific experiment for the Low Energy Demonstration Accelerator (LEDA) facility. To attain this measurement goal, a 52-magnet beam line containing several types of beam diagnostic instrumentation is being installed. The Experimental Physics and Industrial Control System (EPICS) and commercial software applications are presently being integrated to provide a real-time, synchronous data acquisition and control system. This system is comprised of magnet control, vacuum control, motor control, data acquisition, and data analysis. Unique requirements led to the development and integration of customized software and hardware. EPICS real-time databases, Interactive Data Language (IDL) programs, LabVIEW Virtual Instruments (VI), and State Notation Language (SNL) sequences are hosted on VXI, PC, and UNIX-based platforms which interact using the EPICS Channel Access (CA) communication protocol. Acquisition and control hardware technology ranges from DSP-based diagnostic instrumentation to the PLC-controlled vacuum system. This paper describes the control system hardware and software design, and implementation.

## 1 INTRODUCTION

As part of the linac design for the accelerator production of tritium (APT) project the first 10-MeV portion of this 100-mA proton accelerator was assembled at the Los Alamos Neutron Science Center (LANSCE) in 1999 and was in operation for over one year. Now, this Low-Energy Demonstration Accelerator (LEDA) provides the platform for a new experiment: attaining measurements of high-power proton beam-halo formation. For this purpose a 52-magnet beam line has been installed into the LEDA beam line between the Radio Frequency Quadrupoles (RFQ) and the High Energy Beam Transport (HEBT).

LEDA is using the distributed control system based on EPICS [1, 2]. Extensions to the existing control system were developed for controlling the devices listed in Table 1 in section 3.2. This table also shows device locations as defined by the space upstream of the numbered quadrupole magnet.

EPICS is a toolkit for building distributed control systems that originated at Los Alamos and is now developed jointly by a collaboration of over 100 institutions [3]. It is the basis for operator controls interfacing. Specific extensions to EPICS CA communication protocol have been developed to integrate additional controls configuration and visualization options into EPICS. IDL, a commercial visualization tool, has been integrated to provide more complex data processing and visualization options. LabVIEW has been integrated to enable simple, cost effective PC solutions to selective instrumentation control.

Provided below is a summary of the important control system components. Brief descriptions of computer system architecture, hardware, software, and external interfaces are presented. These designs have been, or are being integrated into the control system for LEDA.

## 2 CONTROL SYSTEM STRUCTURE AND NETWORK TOPOLOGY

The Halo Experiment extends the LEDA controls hardware architecture by 3 PC-Input/Output Controllers (IOC)s and 2 VXI-IOCs. The extension is comprised of four principal systems covering Quadrupole Magnet Control, Steering Magnet Control, Diagnostic, and Vacuum System Control.

The control system's communication service is built on a TCP/IP-based network and uses EPICS CA as the primary protocol. Access to the controls network is limited by an Internet firewall for safety and traffic congestion reasons. A second independent local area network has been created to isolate the distributed I/O modules from the control system's network.

## 3 QUADRUPOLE MAGNET CONTROL

The 52-quadrupole-magnet focus/defocus (FODO) lattice provides a platform to create phase space halo formation. The first four quadrupole magnets are each independently powered by a 500A/15V EMI/Alpha-Scientific power supply (Singlet). Depending on how those magnets are adjusted, a match or mismatch of the RFQ output beam to the lattice is created.

The next 48 magnets are powered in groups of 8. Each set of magnets is powered by an Alpha-Scientific 500A/100V Bulk Power Supply with an 8 Channel Shunt Regulator that allows an individual current trim. With certain magnet settings, the development of specific halo formations can be observed along the lattice.

---

[1] This work supported by the Department of Energy under contract W-7405-ENG-36.

### 3.1 Quadrupole Magnet Control Hardware

The quadrupole magnet control subsystem uses as PC-IOC computer an Intel Pentium II 500Mhz equipped with two network cards. The first card establishes the physical connection to the LEDA controls network. The second card creates a local area network for National Instruments (NI) modular distributed I/O system called FieldPoint (FP). It includes analog and digital modules, and intelligent Ethernet network module that connects the I/O modules to the PC-IOC computer.

Using the robust Ethernet networking technology to position intelligent, distributed I/O devices closer to the sensors, or units under test, leads to a significant cost savings and performance improvement. The most obvious benefit to this solution is the savings in signal wiring. Replacing long signal wires with a single, low cost network cable saves significant time and money during installation and testing. Furthermore, distributed I/O systems, such as NI FP, also include special capabilities to improve the reliability and maintainability by using the built-in, onboard diagnostic capabilities.

Table 1. Halo lattice beam line component locations

| Device | Locations (Quadrupole Magnet #) |
|---|---|
| Singlet PS | 1, 2, 3, 4 |
| Bulk PS /Shunts | 5-12, 13-20, 21-28, 37-44, 45-52 |
| Steerers | 4, 6, 15, 17, 26, 28, 36, 38, 47, 49 |
| Fast Valve | 2 |
| Beam Line Valves | 17, 36, 52 |
| Ion Pumps | 6, 10, 15, 21, 28, 34, 40, 44 |
| Ion Gauges | 13, 31, 42 |
| WS/HS | 5, 21, 23, 25, 27, 46, 48, 50, 52 |

### 3.2 Quadrupole Magnet Control Software

The NI LabVIEW-based quadrupole magnet control system that runs on the Windows NT operating system consists of 60 subroutines combined into 10 processes (4 Singlets + 6 Bulk/Shunt Regulator). These independent software processes share CPU cycles based on priority. To ensure proper operation, the control processes for the set-points have highest priority, while the display related processes have lower priorities.

The LabView uses an in-house built ActiveX Automation Server allowing the integration of the LabVIEW system in the EPICS environment by serving values of general interest to Graphical User Interfaces (GUI) that are clients to CA.

Using the EPICS GUI called Display Manager (DM), the control processes drive the power supply values and set-points by interpreting mouse clicks and text entries into commands.

To meet the magnetic field setability specifications, magnet hysteresis must be compensated by ramping the field past the set-point and then reducing the field to the set-point. This led to a control process solution for the Bulk and Singlet power supplies that operates in form of a State Machine where the states are idle, ramp up, soak and ramp down. According to the magnet specification the process control ramps up to a desired set-point, overshoots, stays for a specific time at that overshoot value and ramps back down to the operator desired set-point. Ramp up rate (Amps/step), overshoot value (Amps above desired set-point), soak time (sec), and ramp down rate (Amps/step) are individual settable by the operator. Basic binary operation like AC Voltage On/Off, DC Voltage On/Off, and interlock reset are provided as well.

The display processes are limited to voltage and current read backs, status indication for PS interlocks, and Local/Remote status.

## 4 STEERING MAGNET CONTROL

The task of the beam's 10 steering magnets (horizontal and vertical) is to correct the position of the beam at the end of every set of BPM/steering magnet associated pair that is separated by approximately 10 quadrupole magnets (see Table 1). Since the FODO-lattice period has a phase advance of approximately 80 degrees, each pair of BPMs can detect and each pair of steering magnets can correct the beam's position and angle that might be caused by misaligned quadrupoles.

### 4.1 Steering Magnet Control Hardware

The steering magnet control system uses an Intel Pentium II 450Mhz PC-IOC equipped with one network card and two NI Plug & Play PCI General-Purpose Interface Bus (GPIB/IEEE 488.2) controller cards. This so called two-bus GPIB controller system interfaces 20 KEPCO 20V/5A power supplies, 10 for horizontal and 10 for vertical. All 10 power supplies on each bus are linked together in a linear configuration (daisy-chain). This hardware architecture combines the cost-effectiveness of general-purpose PCs with the standardized and widely used GPIB solution. Due to the limitation of GPIB (max cable length 20m) the PC-IOC is located next to the 2 racks of steering power supplies.

### 4.2 Steering Magnet Control Software

The NI LabVIEW-based steering magnet control system that runs on a Windows NT operating system consists of 80 subroutines incorporated into 1 process. This all-in-one design follows a iterative control sequence between the individual power supplies on both buses: 1) writing to the first power supply on the first bus then writing to the first power supply on the second bus. 2) reading from the first power supply on the first bus and then reading from the first power supply on the second bus. This continues until the 10$^{th}$ power supply has been iterated and then starts over again. This strategy was chosen for possible closed-loop beam correction where the amount of beam spill is critical. Having the described procedure in place reduces possible beam spill by reducing the reaction time of two corresponding steering power supplies to a minimum of ~4ms.

The developed GUI shows the change of current and voltage read backs as the operator changes the current set-points via text entry fields and slider controls. Furthermore, the most important information about the status of the power supplies is read out from the 16-bit status register.

## 5 VACUUM SYSTEM

The devices for the vacuum system for the Halo Experiment beam-line comprise of one Fast Beam-line Isolation Valve, three Beam Line Valves, three Ion Gauges, three Convectron Gauges, and 8 Ion Pumps.

### 5.1 Vacuum System Control Hardware

The vacuum control subsystem operates as a standalone system. That is, all hardware components are local to the Halo Beam-line, hardwired together, network isolated and fully functional, i.e., the subsystem, contains all interlocks and requires no input from remote computer control equipment during normal operation.

Remote accessibility is established through the NI distributed I/O modules called FP (see section 3.2).

### 5.2 Vacuum System Control Software

The DirectSoft PLC is the heart of the vacuum control system and has incorporated all functionality. There is no direct access to the PLC's CPU/memory components during normal operation. Thus, all signals are propagated through the associated PLC input/output modules. A custom ladder logic program containing equipment interlocks, resides and runs (continuous loop) in the volatile memory of the CPU module. Though, the program is lost during a power shut down, the ladder logic program is loaded from the flash memory and started shortly after power is resumed. Its initial state after reboot is a safe mode state in which all devices are turned off.

## 6 WIRE SCANNER/HALO SCRAPER

There are nine Wire Scanner / Halo Scraper (WS/HS) assemblies installed into the LEDA Halo Beam-line [5]. These assemblies contain three measurement devices for each horizontal and vertical axis. Data acquired from these devices is used to provide projected beam distribution information. One wire scanner device measures beam core distribution within +/- 3 rms widths while two halo scraper devices measure the edges of beam distribution outside 2.5 rms widths. These devices are attached to an actuator driven by a stepper motor [6], which drives either the WS or one of the HS into the beam in incremental steps. At each WS step, the amount of secondary electrons (SE) generated is measured and normalized against synchronous beam current data [7]. At each HS step, the proton beam charge is measured and normalized against the overlapping WS data. Together, this provides a complete beam distribution profile.

### 6.1 Control Solutions

Measurement control uses CA Server to communicate between EPICS modules running on Kinetic Systems' HKBaja60 in VXI IOCs, LabVIEW running on a PC IOC, and IDL running on Sun Workstations. The ability of CA Server to provide a communication means between software applications running on different platforms allows the flexibility to choose tools that best suit the specific requirements.

EPICS real-time processing meets the requirement to acquire synchronous beam pulse data at the rates of 1-6 Hz. Data is continuously acquired from beam current monitors, wires, and scrapers by DSPs mounted on the VXI boards into local circular buffers. Data is extracted from these buffers by the EPICS database when triggered.

LabVIEW offers timely and cost effective methods for controlling actuator motion. VIs were implemented to interface with off-the-shelf motor drivers. CA Client runs on the PC to enable communications between LabVIEW and the other control system modules.

IDL provides flexible data processing and visualization options. Operators have the flexibility to process data independently and fine-tune the processing on-line without interrupting beam operations. The results are displayed graphically in plots as they are calculated.

The EPICS' display manager (DM) serves as the interface for operator control. Operators control measurement parameters and view measurement status on a DM GUI.

### 6.2 Implementation

The operator sets parameters and starts the measurement from the WS/HS controls DM GUI. EPICS database process variables are used to pass data and commands between EPICS WS/HS data acquisition, LabVIEW motor control, and IDL data processing/visualization modules. A SNL sequence manages the execution of the separate control modules by setting process variables to known command values. These variables are monitored through CA by associated modules. When the expected command is received by a module, execution is initiated. This flow is illustrated in Figure 1.

For each step in a measurement, the SNL sequence sends motor parameters to LabVIEW via a process variable specifying Z location, horizontal or vertical axis, and stepper motor position. LabVIEW sets a response variable to notify the sequence when the stepper motor has reached the required destination.

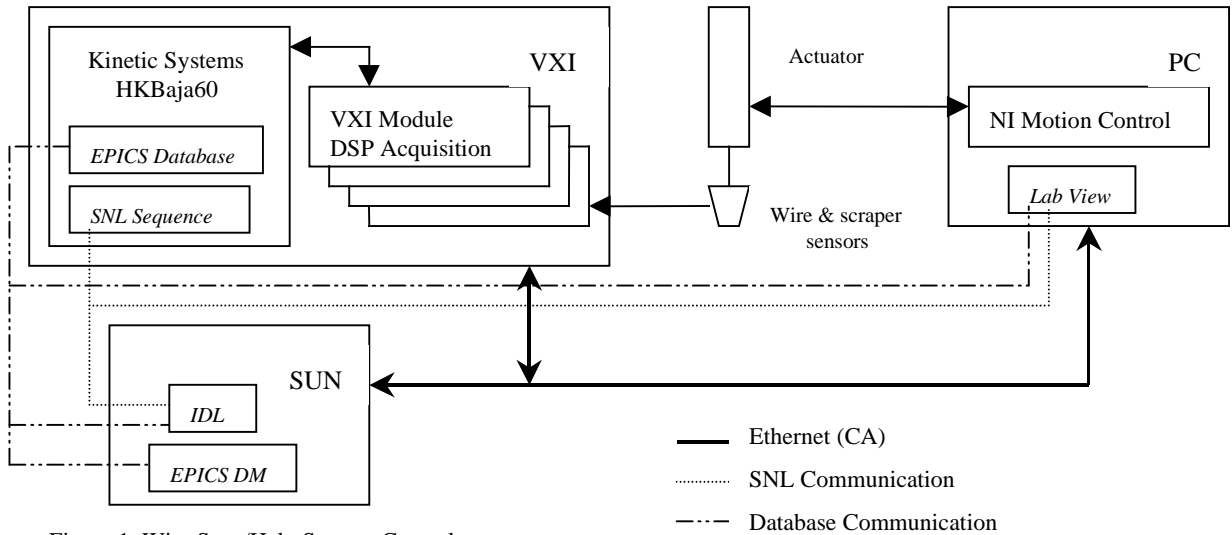

Figure 1: Wire Scan/Halo Scraper Control

Meanwhile, each WS and HS pair has its SE charge waveform signal acquired by its DSP during every beam pulse. When LabVIEW notifies the SNL sequence that the stepper motor is in position, the sequence sets a process variable to trigger the database to upload the active wire or scraper's data along with the associated beam current monitor's data, for normalizing, into waveforms.

This synchronous data is locally processed within the EPICS database and also processed in IDL routines. The local processing completes automatically within the database. These functions produce results of global control system interest such as unit conversion and averaging. Because the EPICS database is loaded at boot time, these functions are fixed, i.e, the majority of parameters cannot be conveniently modified. However, the results are time-stamped, therefore, they can be archived and made available for synchronous data retrieval.

IDL is utilised for specialized and flexible processing and visualization. Functions can be created and/or modified on-line from the control room without affecting beam operations. Furthermore, IDL provides more complex options for viewing data. After data acquisition is complete, the SNL sequence notifies IDL at which time IDL gets all necessary data for processing. The WS or HS data is normalized and plotted against the motor's position producing the beam's profile visually in real time. Operators are able to view the measurement's progress and abort the scan if undesirable results are presented. Processing parameters can be adjusted if desired, and a new measurement started.